\documentclass{article}
\usepackage{graphicx}
\def\bb{\begin{equation}}
\def\ee{\end{equation}}

\def\ve{\varepsilon}

\def\a{\alpha}
\def\b{\beta}
\def\d{\delta}
\def\g{\gamma}

\def\t{\tau}
\def\pt{\partial}
\title{Asymptotic approach for the  rigid condition
of appearance of the oscillations in the solution of the Painleve-2
equation}
\author{O.M.Kiselev,  \\
Institute of Mathematics,\\ Ufa Sci Centre of Russian Acad. of Sci\\
112, Chernyshevsky str., Ufa, 450000, Russia\\ E-mail: ok@
 imat.rb.ru}
\date{February 5,  1999}

\begin{document}
 \maketitle
\begin{abstract}
The asymptotic solution for the Painleve-2 equation with small
parameter is considered. The solution has algebraic behavior before
point $t_*$ and fast oscillating behavior after the point $t_*$. In
the transition layer the behavior of the asymptotic solution is more
complicated. The leading term of the asymptotics satisfies the
Painleve-1 equation and some elliptic equation with constant
coefficients, where the solution of the Painleve-1 equation has poles.
The uniform smooth asymptotics are constructed in the interval,
containing the critical point $t_*$.
\end{abstract}
\vspace{5mm}

\par
In this work a special asymptotic solution on $\ve$ at $\ve\to0$ for
the equation Painleve-2 is constructed:
\bb
\ve^2u'' + 2u^3 + tu = 1.
\label {p2}
\ee
The constructed solution describes a rigid condition of origin of fast
oscillations in some moment $t_*$. At the left, at $t<t_* $, the
asymptotic solution is algebraic, and on the right, at $t>t_*$, this
asymptotics is  fast oscillating. An asymptotic solution in a
transitional layer  (small neighborhood of a critical point $t_*$) is
investigated explicitly. The phase and phase shift of the oscillating
asymptotics are calculated.
\par
The qualitative behaviour of solutions of second-order ordinary
differential equations according to an additional parameter is
explained, for example, in the book \cite{Andr}. In \cite{Andr} the
various types of bifurcations for equilibrium positions of
conservative second-order ordinary differential equations are
described also.
\par
Consider the autonomous equation obtained from (\ref {p2}). If we "freeze"
a value of variable coefficient $t=T$ then we obtain the equation:
\bb
\ve^2V'' + 2V^3 + TV = 1.
\label{el}
\ee
This equation has three solutions which are according to equilibrium
positions of dynamical system for this equation, if the parameter $T$
is less than some bifurcation value $t_*$. Two of this equilibrium
positions are stable and one position is unstable. If the parameter
$T>t_*$ then the dynamical system has only one equilibrium position.
This position is stable. The other solutions of the equation are
described by some elliptic function. The construction of asymptotics
of a solution for this differential equation near to a bifurcation
point of a parameter $T$ is reduced to a research  of the elliptic
function asymptotics.
\par
The equation (\ref{p2}) is nonautonomous and therefore the existence
"an equilibrium positions" depends on the variable $t$. In a critical
point $t=t_*$ the bifurcation of a "saddle-node" type happens and at
$t>t_*$ one of stable "an equilibrium position" disappears. The
bifurcation of such type leads to the rigid loss of a stability \cite
{Arn}. The "saddle - node" is the elementary bifurcation for the
second-order ordinary differential equations. The solution for the
equation (\ref {p2}) in the part of the neighborhood for the point of
the "saddle-node" bifurcation is described by the equation Painleve-1.
The solution of the Painleve-1 equation has poles. But the asymptotic
solution of the equation (\ref{p2}) have not the poles. Therefore,
near the poles of the solution for the Painleve-1 equation we can't
use the Painleve-1 equation. In such areas of the variable $t$ the
asymptotic solution is described by some autonomous nonlinear
equation.
\par
The next in complexity bifurcation is the doubling originating in the
equation Painleve-2. Asymptotics with respect to a small parameter of
the solutions for the equation Painleve-2 with zero in the right hand
side in  the equation (\ref{p2}) were investigated in works
\cite{Hab}, \cite{Kar-Per}. In this case solution in an interior layer
near to a bifurcation value of the parameter $t_*$ is determined by
the equation Painleve-2, but already without a small parameter; and
the problem, generally speaking, does not become simpler.
\par
The structure of the constructed asymptotic solution is various in
various ranges of values of a parameter $t$. The asymptotic solutions,
suitable at $t<t_*$ and in a small neighborhood $t_*$, are constructed
using the direct method of the theory of perturbations \cite{Nife}.
The fast oscillating asymptotics at $t>t_*$ is constructed by the
Krylov-Bogolyubov's method \cite{K-B}, explained for nonlinear
second-kind ordinary equations in the work by G.E. Kuzmak: \cite{Kuz}
and justified by M.V. Fedoryuk in \cite{Fed}. The various types of
asymptotics are matched by the matching method for the asymptotic
expansions \cite{I}.
\par
The asymptotics of the solutions for the Painleve equations with a
leading term as an elliptic function with the modulated parameters
were studied, for example, in works \cite{B} -\cite{V}. The
neighborhood of the degenerated point of the elliptic anzatz in these
works was not investigated. The scaling limit passages of the equation
Painleve-2 to the equation Painleve-1 and to the  sine-elliptic
equation \cite{B}, \cite{Kap} are known also. However, the asymptotic
solutions constructed by this way, for example, in \cite{Kap} are
non-uniform on the variable $t$. The qualitative analysis for the
relation between the algebraic and fast oscillating asymptotic
solutions of the equation (\ref{p2})was done in the work
\cite{OK-BIS}.
\par
Here the equation Painleve-2 is considered as a model example for
study of a connection of various types of movements circumscribed by
one solution of the nonlinear equation. When choosing the equation we
take in account only its simply form and the statement about
boundedness of a solutions  for this equation \cite{Ince}. In this
work the properties of an integrability of the equation (\ref{p2}) by
the inverse scattering transform \cite{F-N} are not used anywhere.
\par
In this work we do not prove any theorems about the justification of
the constructed asymptotics for the equation (\ref {p2}). However, in
that part of researched area with respect to variable $t$ , where the
formal asymptotic solution is oscillates rapidly, the justification is
based on the results of the work \cite{Fed}.

\section {Naive statement of the problem}

\par
Let's consider a cubic equation obtained at a rejection of the term
with the small parameter in the equation (\ref {p2}):
\bb
2u^3 + tu = 1.
\label{kub}
\ee
This equation has three real roots $u_1(t)<u_2(t)<u_3(t)$ at $T<t_*$
and one real root at $t>t_*$; if $t=t_*$, then the roots $u_1(t)$ and
$u_2(t)$ stick together $u_1(t_*)=u_2(t_*)=u_*$. The values $u_*$ and
$t_*$ are easy for obtaining, by solving of the equations:
$$
2u_*^3 + t_*u_*-1 = 0, \quad 6u_*^2 + t_*=0.
$$
\par
At $t<t_*$ it is possible to construct an asymptotic solution of the
equation (\ref {p2}) by taking as a leading term of asymptotics any of
the solutions for the equation (\ref {kub}). Thus the asymptotic
solution with the leading term $u_1(t)$ or $u_3(t)$ is stable and if
the leading term is $u_2(t)$ then the asymptotics is unstable. It
follows from the analysis of the linearized equation (\ref {p2}):
$$
\ve^2v'' + (6u_j^2(t)+t)v = 0, \quad j = 1,2,3.
$$
\par
The asymptotic solution with the leading term $u_3(t)$ keeps its
structure in the neighborhood of the bifurcation  point $t_*$ and
after this point: at $t>t_*$. Unlike it, the structure of the
asymptotic solution with the leading term $u_1(t)$ varies sufficiently
at passage through the bifurcation point $t_*$.
\par
Our problem is to construct a smooth formal asymptotic solution
of the equation (\ref{p2}) on a segment $[t_*-a, t_*+a],
\quad a=const>0$ with the leading term $u_1(t)$ at $t<t_*$.

\section {The main results}

\par
In this work the smooth formal asymptotic solution of the equation
(\ref{p2}) on a segment $L=[t_*-a, t_*+a], \quad a=const>0$ is
constructed. This solution has a various structure in the different
areas of the segment $L$.
\par
At $(t_*-t)\gg\ve^{4/5}$ the formal asymptotic solution has the form
\bb
u(t,\ve)=\stackrel{0}{u}(t)+\ve^2\stackrel{1}{u}(t)+
O(\ve^4(t-t_*)^{-9/2}).
\label{exp10}
\ee
The leading term is least of the roots of the cubic equation (\ref
{kub}). The gauge sequence in this case is following:
$\ve^{2n},\,n=0,1,2,\dots$. The coefficients of the asymptotic
expansion are the algebraic functions with respect to the parameter
$t$.
\par
At $|t-t_*|\ll1$ the asymptotics is defined by two various
types of the asymptotic expansions. One of them has the form
\bb
u(t,\ve)=u_* + \ve^{2/5}\stackrel{0}{v}(\t)+O(\ve^{4/5}\t)
+O((\t-\t_k)^{-4}\ve^{4/5}).
\label {exp20}
\ee
The function $\stackrel{0}{v}(\t)$ is defined as the solution of the
equation Painleve-1:
$$
{d\stackrel{0}{v}(\t)\over d\t}+6u_*\stackrel{0}{v}\!\!^2+u_*\t=0,
$$
with the pure algebraic asymptotics at $ \t\to-\infty $:
$$
\stackrel{0}v(t)=-\sqrt{-\t\over6}+O(\t^{-2}).
$$
Here the variable $\t$ is define by the formula
$\t=(t-t_*)\ve^{-4/5}$.
\par
The function $\stackrel{0}{v}(\t)$ has poles in some points $\t_k>0$
\cite{H-S}. Outside of these poles the expansion (\ref{exp20}) is
suitable.
\par
Near the poles of the function $\stackrel{0}{v}(\t)$ in the areas
$|\t-\t_k||\t_ k|^{1/5}\ll1$ the asymptotic solution of the equation
(\ref{p2}) has an another structure:
\bb
u(t,\ve)=u_*+\stackrel{0}{w}(\theta)+O(\ve^{4/5}\theta^2|\t_k|),
\label{exp30}
\ee
where $\theta=(\t-\t_k)\ve^{-1/5}$. The function
$\stackrel{0}{w}(\theta)$ is defined by the formula:
$$
\stackrel{0}{w}(\theta)={-16u_*\over4 + 16u_*^2\theta^2}.
$$
\par
At $(t-t_*)\ve^{-4/5}\gg1$ the asymptotics have the fast oscillating
behavior:
\bb
u(t,\ve)=\stackrel{0}{U}(t_1,t)+O(\ve).
\label{exp40}
\ee
The leading term of the asymptotic solution satisfies the equation
$$
(S')^2(\stackrel{0}{U}_{t_1})^2=-\stackrel{0}{U}^4-
t\stackrel{0}{U}^2+2\stackrel{0}{U}+E(t),
$$
where $t_1 = S(t)/\ve$. The function $E(t)$ is defined by the equation
$$
\int_{\b(t)}^{\a(t)}\sqrt{-x^4-tx^2+2x+E(t)}dx=\pi,
$$
where $\a(t)$ and $\b(t)$ were solutions of the  equation $-x^4-tx^2 +
2x + E(t)=0$.
\par
The phase function $S(t)$ is the solution of the Cauchy problem:
$$
T=S'\sqrt{2}\int_{\b(t)}^{\a(t)}{dx\over\sqrt{-x^4-tx^2+2x+E(t)}},
\quad S|_{t=t_*}=0,
$$
where $T$ is some constant defined in the formula (\ref{T}).
\par
{\bf Note about the matching of the asymptotics.} The  areas of
usefulness for the asymptotic expansions (\ref{exp10}) and
(\ref{exp20}), (\ref{exp20}) and (\ref{exp30}), and also
(\ref{exp20}), (\ref{exp30}) and (\ref{exp40}) are intersected. It
allows to matching this asymptotics.
\par
{ \bf Note about rigorously.} In areas, where the asymptotic
expansions (\ref{exp10}) and (\ref{exp30}) are suitable, it is
possible to construct the full asymptotic expansions with respect to
parameter $\ve$ and, probably,  to justify them. The justification of
the fast oscillating asymptotics (\ref{exp40}) follows from the work
\cite{Fed}. A more complicated situation is with the expansion
(\ref{exp20}). The problem about the construction of the full
asymptotic expansion of the type (\ref{exp20}) is tightly associated
with the problem about the perturbation of the equation the
Painleve-1. As far as it is known, the problems about the perturbation
of the Painleve equations are investigated insufficiently. The
detailed research of the perturbed equation Painleve-1 is possible
probably on the basis of the monodromy-preserving deformations method
\cite{F-N}, but it goes out for the framework of the present work.

\section {The qualitative analysis of the problem and of the results}

\par
We assume, that the constructed asymptotic solution of the equation
(\ref{p2}) is an asymptotics of some true solution. The equation
(\ref{p2}) is nonautonomous, for the qualitative analysis of its
solutions at $t<t_*, \quad t\sim t_*, \quad t>T_*$ we shall take
advantage the equation with the "frozen" value of the coefficient $t$
in the equation (\ref{p2}) in some point $t=T$. That is, we shall
consider the equation (\ref {el}). Integrate it once with respect to
$T$, in the result we shall get:
$$
\ve^2 (V')^2=-V^4-TV+2V+E, \quad E=const.
$$
Depending on the value of the parameter $T$ the potential for the
equation \ref{el}) has one of graphs:

\includegraphics[width=100mm,angle=0]{osc1-1.ps}\\
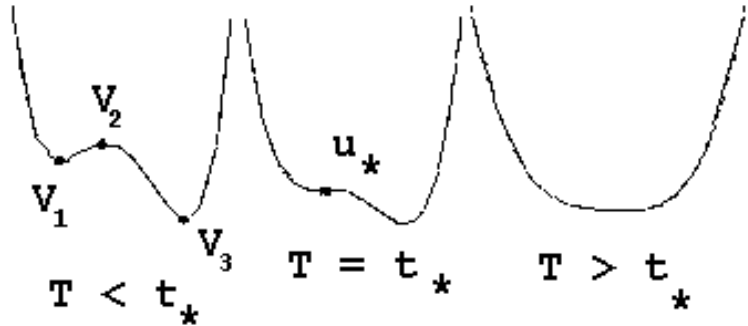
\centerline{Fig.1}

At $T<t_*$ the points $V_1$ and $V_3$ are the point of a stable
equilibrium, and $V_2$ is the unstable equilibrium. At $T=t_*$ the
points $V_1$ and $V_2$ stick together and $u_*$ is the point of
unstable equilibrium. At $T>t_*$ there is only one point of the
equilibrium. Thus, at $T<t_*$ on the phase plane of the equation
(\ref{el}) there are two centers and one saddle; at $T=t_*$ one of
centers sticks together to the saddle; at $T>t_*$ there is only one
center. It is the bifurcation of the "saddle - node" type.
\par
Let's explain the connection of figure 1 with the equation (\ref{p2}).
It is easy to see, that at $t=T<t_*$ the value of the leading term of
the algebraic asymptotic solution (the function $u_1(t)$) coincides
with $V_1$. At $T=t_*$ the points $V_1$ and $V_2$ stick together and
the solution $u(t,\ve)$ begins to change with an energy which is close
to $E_*=u_*^4+t_*u_*^2-2u_*$. On the greater part of the trajectory
the movement is fast and the characteristic variable is
$\theta\sim\ve^{-1}t$. On the slanting part (near to the point $u_*$)
the characteristic variable is decelerated: $\t=(t-t_*)\ve^{-4/5}$. At
$t>t_*$ the movement is close to periodic and characteristic fast
variable is $t_1=S(t)/\ve$.
\par
The results of the numerical resolving of the equation (\ref{p2}) by
the Runge-Kutta method for the Cauchy problem with the entry
conditions corresponding to the algebraic asymptotics at $t=-7$ and
$\ve=0.2$ give the similar solution.

\includegraphics[width=100mm,angle=0]{osc1-2.ps}\\

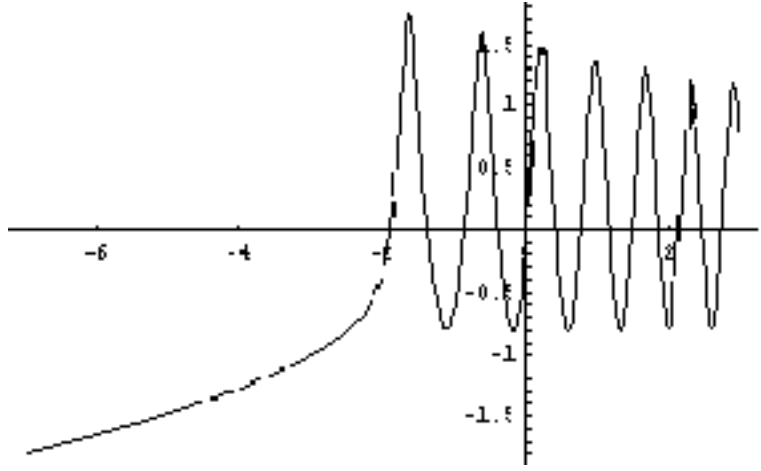
\centerline{Fig.2}

\section{The exterior algebraic asymptotics}

\par
The algebraic asymptotic solution of the equation (\ref{p2}), which is
suitable at $t<t_*$, is constructed here and its asymptotics is
investigated at $ t\to t_*-0$.
\par
For the algebraic asymptotic solution of the equation (\ref{p2}) we
search as:
\bb
u(t,\ve)=\stackrel{0}{u}(t)+\ve^2\stackrel{1}{u}(t)+\ve^4\stackrel{2}{u}(t)
+ \dots.
\label{exp1}
\ee
\par
Let's formulate the result of this section. The formal asymptotic solution
(\ref{exp1}), where $\stackrel{0}{u}(t)$ is least of the solutions of
the equation (\ref{kub}), is suitable at $ (t_*-t)\ve^{-4/5}\gg1$.
\par
Begin to obtain the coefficients of the asymptotics (\ref{exp1}).
Substitute the anzatz (\ref{exp1}) to the equation (\ref{p2}). Let's
equate the coefficients at identical powers of $\ve$. In the result we
obtain the recurrence sequence of the formulas for the definition
$\stackrel{k}{u}(t),\quad k=0,1,2,\dots$.
$$
2\stackrel{0}{u}\!^3(t)+t\stackrel{0}{u}(t)=1,\quad
(6\stackrel{0}{u}\!^2(t)+t)\stackrel{1}{u}(t)=-\stackrel{0}{u}\!''(T),
$$
$$
(6\stackrel{0}{u}\!^2(t)+t)\stackrel{2}{u}(t)=
-6\stackrel{0}{u}(t)\stackrel{1}{u}\!^2(t)-\stackrel{1}{u}''(t).
$$
\par
The cubic equation for $\stackrel{0}{u}(t)$ at $t<t_* $ has three real
roots $u_1(t)<u_2(t)<u_3(t)$. As the leading term of asymptotic
expansion (\ref{exp1}) we choose $u_1(t)$. Let's calculate the second
derivative of $\stackrel{0}{u}(t)$ and  express $\stackrel{1}{u}(t)$
through $\stackrel{0}{u}(t)$:
\bb
\stackrel{1}{u}(t)=
{12\stackrel{0}{u}\!^3(t)\over(6\stackrel{0}{u}\!^2(t)+t)^4}
-{2\stackrel{0}{u}(t)\over(6\stackrel{0}{u}\!^2(t)+t)^ 3}.
\label{u1}
\ee
\par
It is easy to get the expressions for the following terms of the
asymptotics (\ref{exp1}). In an explicit form they are not reduced
here, however, it is important to notice, that the power of the
denominator $(6\stackrel{0}{u}\!^2(t)+t)$ in the coefficients of the
asymptotics grows with each next step. At
$(6\stackrel{0}{u}\!^2(t)+t)\to0$ $n$-th term of the asymptotic
expansion has the form
\bb
\stackrel{n}{u}(t)=O\big((6\stackrel{0}{u}\!^2(t)+t)^{-5n+1}\big).
\label{un}
\ee
\par
Let's write out the asymptotics of the asymptotic expansion (\ref{exp1})
at $t\to t_*$. For this purpose we shall calculate the asymptotics of the
expression $(6\stackrel{0}{u}\!^2(t)+t)$:
$$
(6\stackrel{0}{u}\!^2(t)+t)|_{t\to t_*}=-2u_*\sqrt{6}\sqrt{t_*-t}
+{2\over3}(t_*-t)+{-5\over9\sqrt{6}u_*}(t-t_*)^{3/2}+O((t_*-t)^2).
$$
Using this formula and (\ref{exp1}), (\ref{u1}), we get:
$$
u(t,\ve)=u_*-{1\over\sqrt6}\sqrt{t_*-t}+{1\over18u_*}(t_*-t)+
$$
$$
+
\ve^2\bigg[-{1\over3\,2^{10/3}}(t_*-t)^{-2}-O((t_*-t)^{-3/2})\bigg]
+O\bigg(\ve^4(t_*-t)^{-9/2}\bigg)+O((t_*-t)^{3/2}).
$$
The area of usefulness for this expansion at $t\to t_*-0 $ is
determined from the relation
$\ve^2\stackrel{n+1}{u}(t)/\stackrel{n}{u}(t)\ll1$. It follows from
the formula (\ref{un}), that the expansion (\ref{exp1}) is suitable at
$(t_*-t)\ve^{-4/5}\gg1$.

\section {The interior asymptotics}

\par
In this section the asymptotic expansions which are suitable in the
small neighborhood of a point $t_*$ are constructed. Following the
terminology of the matching method \cite {I}, they are called "the
interior asymptotic expansions".

\subsection {First interior expansion}

\par
In the neighborhood of the point $t_*$ we make the stretch, which are
dictated by the asymptotics at $t\to t_*$ of the exterior expansion
(\ref{exp1}):
$$
(u-u_*)=\ve^{2/5}v, \quad (t-t_*)=\ve^{4/5}\t.
$$
As result the equation (\ref{p2}) we write as
\bb
{d^2v\over d\t^2}+6u_*v^2 + u_*\t=-\ve^{2/5}(\t v+2v^3).
\label{p1}
\ee
\par
In the limit at $\ve\to0$ we obtain the equation Painleve-1. This
asymptotic reduction is known as one of the scaling limits for the
Painleve-2 equation \cite {Kap}.
\par
The asymptotic solution of the equation (\ref{p1}) we build as:
\bb
v(\t,\ve)=\stackrel{0}{v}(\t)+\ve^{2/5}\stackrel{1}{v}(\t),
\label{exp2}
\ee
where the function  $\stackrel{0}{v}(\t)$ is the solution of the
Painleve-1 equation.
\par
Here it is shown, that the asymptotic solution (\ref{exp2}) is
suitable in the neighborhood of infinity (at $\t\ll\ve^{-4/5}$) and in
the neighborhood of the poles for the function $\stackrel{0}{v}(\t)$:
$ (t-\t_k)\ve^{-1/5}\gg1 $.
\par
The coefficients of the asymptotics are calculated from the matching
condition for the asymptotic expansion (\ref{exp1}) at
$t\to t_*$ and the expansion (\ref{exp2}) at $\t\to-\infty$. In particular,
$\stackrel{0}{v}(\t)$ has the algebraic asymptotics:
\bb
\stackrel{0}{v}(\t)|_{\t\to-\infty}={-1\over\sqrt6}\sqrt{-\t}
+O((-\t)^{-2}).
\label{asp11}
\ee
\par
In the book \cite{G-L} it is shown, that there is the solution of the
Painleve-1 equation with the asymptotics (\ref{asp11}). In the work
\cite{H-S} it is proved, that the solution of the Painleve-1 equation
with the asymptotics (\ref{asp11}) has not poles at $\t\le0$. The data
of a monodromy for the solution of the Painleve-1 equation with the
asymptotics (\ref{asp11}) are calculated in the work \cite{Kap2}.
\par
The first correction in the asymptotics (\ref{exp2}) satisfies the
equation
\bb
{d^2\stackrel{1}{v}\over d\t^2}+12u_*\stackrel{0}{v}\stackrel{1}{v}=
-\t\stackrel{0}{v}-2\stackrel{0}{v}\!^3.
\label{lp1}
\ee
The asymptotics of the solution for this equation at $\t\to-\infty$
has the form
$$
\stackrel{1}{v}(\t)={\t\over18u_*}+O ((-\t)^{-3/2}).
$$
\par
The requirement of fitness for the asymptotics is
$\ve^{2/5}\stackrel{1}{v}/\stackrel{0}{v}\ll1$. It reduces to the
condition $(-\t)\ll\ve^{-4/5}$.
\par
On the positive semiaxis the function $\stackrel{0}{v}(\t)$ has the
poles. Let's denote this poles by $\t_k$. In the neighborhood of the
pole $\t=\t_k$ the function $\stackrel{0}{v}(\t)$ is defined by the
converging power series \cite{G-L}
\bb
\stackrel{0}{v}(\t)={-1\over u_*(\t-\t_k)^2}+
{\t_k u_*\over10}(\t-\t_k)^2{u_*\over6}(\t-\t_k)^3+c_4(\t-\t_k)^4+
O((\t-\t_k)^5).
\label{asp12}
\ee
The constants $\t_k$ and $c_4$ are the parameters of this solution. In
the review \cite{Kit} it is marked, that the problem on the connection
between the asymptotics of this solution at infinity and the constants
$\t_k$ and $c_4$ is not investigated yet. The points of the poles
$\t_k$ and appropriate constants $c_4$ can be obtained with the help
of the numerical calculation using the given asymptotics at infinity
(\ref{asp11}). The asymptotics $\stackrel{1}{v}$ at $\t\to\t_k$ has
the form
\bb
\stackrel{1}{v}={-1\over (\t-\t_k)^4}+{19\t_k\over10u_*}
-{5\over24u_*^2}(\t-\t_k)+O((\t-\t_k)^2).
\label{aslp1}
\ee
\par
Using the asymptotics (\ref{asp12}) and (\ref{aslp1}) it is easy to
see, that the asymptotic expansion (\ref{exp2}) is suitable at
$$
|\t-\t_k|\gg\ve^{1/5}.
$$

\subsection {Second interior expansion}

\par
For the construction of the uniform asymptotics in the neighborhood of
the pole of the function $\stackrel{0}{v}$ it is necessary  to make
one more stretching of the independent variable and the function:
$$
(\t-\t_k)=\ve^{1/5}\theta, \quad\ve^{-2/5}v=w.
$$
For function $w$ we obtain the equation:
\bb
{d^2 w\over d\theta^2}+6u_*w^2+2w^3=-\ve^{4/5}\t_k(u_*+w)-\ve\theta(u_*+w).
\label{de3}
\ee
\par
We search the asymptotic solution of this equation as
\bb
w(\theta,\ve)=\stackrel{0}{w}+\ve^{4/5}\stackrel{1}{w}+\ve\stackrel{2}{w}
+\dots.
\label{exp3}
\ee
\par
It is shown here, that the asymptotic expansion (\ref{exp3}) is the
formal asymptotic solution of the equation (\ref{de3}) at
$|\theta\t_k^{1/5}|\ll\ve^{1/5}$.
\par
The solution of the equation for the leading term of the asymptotics
(\ref{exp3}) is not uniquely defined determined from the asymptotics
at $\t\to\t_k$ of the asymptotic expansion (\ref{exp2}), which is
exterior in relation to (\ref{exp3}). This solution has the form
\bb
\stackrel{0}{w}(\theta)={-16u_*\over4+16u_*^2\theta^2}.
\label{sol}
\ee
\par
The corrections in the expansion (\ref {exp3}) satisfy the linearized
equations
$$
{d^2\stackrel{1}{w}\over d\theta^2}+
(12u_*\stackrel{0}{w}+6\stackrel{0}{w}\!^2)\stackrel{1}{w}=
-\t_k(u_*+\stackrel{0}{w});
$$
$$
{d^2\stackrel{2}{w}\over d\theta^2}+
(12u_*\stackrel{0}{w}+6\stackrel{0}{w}\!^2)\stackrel{2}{w}=
\theta(u_*+\stackrel{0}{w}).
$$
\par
The expression for $\stackrel{0}{w}$ can be used to obtain two
linearly independent solutions of the homogeneous equation for the
corrections:
$$
w_1={8\theta\over(1+4u_*^2\theta^2)^2},
$$
$$
w_2=[-{1\over8}+2u_*^2\theta^2-u_*\theta^4+
{2\over5}\theta^6+{2\over7}\theta^8]{1\over(1+4u_*^2\theta^2)^2}.
$$
\par
Using these solutions of the homogeneous equation, it is easy to get
the solutions of the nonhomogeneous equations for the corrections. The
asymptotics of the corrections  at $\theta\to\infty$ has the form:
$$
\stackrel{1}{w}={\t_ku_*\over10}\theta^2
({\t_ku_*^2\over10}+{1\over12u_*})+O(\theta^{-2});
$$
$$
\stackrel{2}{w}={u_*\over6}\theta^3+{1\over u_*}\theta+O(\theta^{-5}).
$$
\par
Using the asymptotics for the corrections and the leading term, we
obtain, that the expansion (\ref{exp3}) is suitable at
$|\theta\t_k^{1/5}|\ll\ve^{-1/5}$. On the other hand, the expansion
(\ref{exp2}) is suitable at $|\theta|\gg1$. Hence, the areas of the
usefulness for the expansions (\ref{exp2}) and (\ref{exp3}) are
intersected at realization of the condition $|\t_k|\ll\ve^{-1}$. If we
take into account also the requirement of fitness of the asymptotic
expansion (\ref{exp3}), then get the restriction
$|\t_k|\ll\ve^{-4/5}$. From this restriction it follows, that in this
section the formal asymptotic expansions suitable at $|t-t_*|\ll1$ are
constructed.

\subsection {The asymptotics of the interior expansions at $\t\to\infty$}

\par
The asymptotics of the solution of the equation Painleve-1 at
$\t\to\infty$ can be obtained with the help of the
monodromy-preserving method \cite{F-N}. If the monodromy data are
known, then the defining solution is not uniquely defined. In the
correspondence with \cite{Kap2}, in our case the monodromy data are
constants $s_2$ and $s_3$ and these constants are equal to zero. The
asymptotics of the function $\stackrel{0}{v}(\t)$ outside of the poles
has the form \cite{Kit}
\bb
\stackrel{0}{v}=\sqrt{\t}\rho(\phi(\t))+O(\t^{-\g }),
\label{asP1}
\ee
Where $\g>0$ is some constant, the function $\rho(x)$ satisfies the
equation
$$
\rho''+6u_*\rho^2+u_*=0,
$$
The phase function $\phi(\t)={4\over5}(\t)^{5/4}$.
\par
It is important to note, that in the formula (\ref{asP1}) the shift of
the phase function $\phi(\t)$ is equal to zero. The asymptotics of
$\stackrel{0}{v}(\t)$ is determined by the  Weierstrass elliptic
function, through which the function  $\rho$ is expressed in
\cite{Kit}.
\par
In the second interior expansion the asymptotics at $\t\to\infty$
corresponds to the asymptotics at $\t_k\to\infty$. It is easy to see,
that in this case the first correction grows. This growth limits the
value $\t_k$, at which the asymptotics (\ref{exp3}) is correct:
$|\t_k|\ll\ve^{-4/5}$.

\section {Fast oscillating asymptotics}

\subsection {The Kuzmak's approximation}
\par
The formulas obtained in the work \cite{Kuz} for the construction of
the fast oscillating asymptotic solution of the second-order ordinary
differential equation by the Krylov-Bogolyubov's method are obtained
here and the parameters of this solution are specified, at which it is
degenerated in the point $t=t_*$.
\par
Let's pass to construction of the oscillating asymptotic solution.
Following \cite{Fed} we search for it as
\bb
u(t,\ve)=\stackrel{0}{U}(t_1,t)+\ve\stackrel{1}{U}(t_1,t) + \dots.
\label{exp4}
\ee
As the argument $t_1$ we use expression $S(t)/\ve$, where $S(t)$ is
known function. The equations for the leading term and  the first
correction of the asymptotics (\ref{exp4}) look like:
\bb
(S')^2\pt_{t_1}^2\stackrel{0}{U}+2\stackrel{0}{U}\!^3+\stackrel{0}{U}t=1,
\label{eq-2U}
\ee
$$
(S')^2\pt_{t_1}^2\stackrel{1}{U}+(6\stackrel{0}{U}\!^2+t)\stackrel{1}{U}=
-2S'\pt^2_{tt_1}\stackrel{0}{U}-S''\pt_{t_1}\stackrel{0}{U}.
$$
Integrate once with respect to $t_1$ the equation for
$\stackrel{0}{U}$ and we obtain in the result:
\bb
(S')^2(\pt_{t_1}\stackrel{0}{U})^2=-\stackrel{0}{U}\!^4t\stackrel{0}{U}\!^2
+2\stackrel{0}{U}+E(t),
\label{eq-U}
\ee
where $E(t)$ is the "constant of integration".
\par
In \cite{Kuz} it is shown, that the condition of periodicity with
respect to the parameter $t_1$ for the function $\stackrel{1}{U}$
reduces to the equation for the function $S(t)$:
$$
S'\int_0^T\big[\pt_{t_1}\stackrel{0}{U}(t_1,t)\big]^2dt_1=c_0.
$$
Here $T$ is the period of the oscillations, $c_0$ is the constant.
Using the explicit expression for the derivative with respect to $t_1$
we can present this formula in a little bit other form:
\bb
2\int_{\b(t)}^{\a(t)}\sqrt{-x^4-tx^2+2x+E(t)}dx=c_0,
\label{eq-S}
\ee
where $\a(t)$ and $\b(t)$ are the solutions of the equation
$-x^4-tx^2+2x+E(t)=0$. The function $E(t)$ is not defined here. Its
connection with the phase of the fast oscillations $S(t)$ is given by
the formula \cite{Fed}:
\bb
T=\sqrt{2}S'\int_{\b(t)}^{\a(t)}{dx\over\sqrt{-x^4-tx^2+2x+E(t)}}.
\label{eq-E}
\ee
{\bf Note.} A.N. Belogrudov has mentioned, that the integral in the
left part (\ref{eq-S}) is hypergeometric function satisfying a system
of equations in partial derivatives with respect to parameters $\a$
and $\b$ \cite{GR}.
\par
The equations (\ref{eq-U})-(\ref{eq-E}) define to an accuracy of some
constant $c_0$ the leading term of the asymptotics (\ref{exp4}). Let's
remark, that we build the asymptotic solution of the equation
(\ref{p2}) at $t>t_*$. Thus the polynomial of the fourth power on
$\stackrel{0}{U}$ in the right hand side of the equation (\ref{eq-U})
can have no more two various real roots $\a(t)$ and $\b(t)$. Hence
this polynomial can be submitted as:
$$
F(x,t)=(\a(t)-x)(x-\b(t))\Big((x-m(t))^2+n^2(t)\Big).
$$
\par
In the point $t=t_*$ the curve on Figure 1 has the inflection point.
The degeneration of the elliptic integral at $t=t_*$ corresponds to
the case $m(t_*)=\b(t_*)=u_*$ and $n(t_*)=0$,  when one of the roots
of the polynomial corresponds to the value of the polynomial in the
inflection point. For this case it is easy to calculate the constant
in the right hand side of the equation (\ref{eq-S}) $c_0=\pi$ and the
value of the parameter $E(t_*)=E_*$.

\subsection {Degeneration of the fast oscillating asymptotics}

\par
We shown here, that the formal asymptotic solution (\ref{exp4})
obtained in the previous subsection is suitable at
$(t-t_*)\ve^{-4/5}\gg1$. The matching by this asymptotics with the
interior asymptotics (\ref{exp2}) and (\ref{exp3}) is carried out.
>From the matching condition for the phase function $S(t)|_{t=t_*}=0$
is obtained.
\par
The oscillating solution is degenerated at $t\to t_*+0$. Let's
construct the asymptotics of this solution in the neighborhood of the
degeneration point. For this purpose we calculate the asymptotics of
the phase function $S(t)$ and the function $E(t)$. Let's write the
equation (\ref{eq-S}) as:
\bb
\int^{\a}_{\b}\sqrt{(\a-x)(x-\b)[(x-m)^2+n^2]}dx=\pi,
\label{eq-E1}
\ee
where $\a,\b,m,n$ are real functions at $t\ge t_*$. These functions
satisfy the Vieta equations:
\begin {eqnarray}
\a + \b + 2m = 0, \nonumber \\
m^2 + n^2 + \a\b + 2m(\a + \b) = t, \nonumber \\
( \a + \b) (m^2 + n^2) + 2m\a\b = 2, \nonumber \\
\a\b (m^2 + n^2)=-E.
\label{vieta}
\end {eqnarray}
\par
The equation (\ref{eq-E}) and three equation from (\ref{vieta}) define
the dependency $ \a, \b, m, n $ with respect to the parameter $t$. The
last equation in (\ref{vieta}) defines the function $E(t)$. Let's make
changes of variables: $E=E_*+g_1, \quad t=t_*+\eta,
\quad m=m_*+m_1.$ After the simple transformations of the equations
(\ref{vieta}) we get:
\begin{eqnarray}
2m_*[6m_1^2-2n^2+\eta]+[2m^2_1-2n^2+\eta]2m_1=0,\nonumber\\
m_*^2(12m_1^2-4n^2+\eta)+2m_*m_1(6m_1^2-2n^2+\eta)+\nonumber\\
(3m^2_1-n^2+\eta)(m^2+n^2)=-g_1.
\label{v2}
\end{eqnarray}
Construct the solution of this system at $t\to t_*+0$ as:
\begin{eqnarray*}
m_1=\mu\sqrt{\eta}+O(\eta),\\
n=\nu_1\sqrt{\eta}+O(\eta),\\
g_1=\g_1\eta+O(\eta^{3/2}).
\end{eqnarray*}
\par
Let's substitute these expressions in (\ref{v2}), equate the
coefficients at the identical powers of $\eta$. In the result we
obtain:
$$
6\mu_1^2-2\nu_1^2=-1, \quad
\g_1=m_*^2.
$$
\par
To define the constants $\mu_1$ and $\nu_1$ it is necessary to
construct the asymptotics at $\eta\to +0$ of the left hand side of the
equation (\ref{eq-E1}). The asymptotics of the outside the integral
coefficient in the equation (\ref{eq-E1}) has the form
\bb
(\a-\b)^3=64|m|^3\big[1-{3\over2}{\mu_1\sqrt{-\eta}\over m_*}+{3\over2}
{\nu_1^2-\mu_1^2-1\over4m_*^2}\eta+O(\eta^{3/2})\big].
\label{a-b}
\ee
\par
The integral in the equation (\ref{eq-E1}) is presented as
$$
I(k,\d)=\int_0^1\sqrt{(1-z)z}\sqrt{(z-k\d)^2+\d^2},
$$
where
\bb
{m-\b\over\a-\b}=k\d, \quad \d={n^2\over(\a-\b)^2},
\label{k1}
\ee
The value of the constant $k$ will be defined from an asymptotics
below.
\par
The asymptotics of an integral $I(k,\d)$ at $\d\to0$ has the form
\bb
I(k,\d)={\pi\over16}-k\d{\pi\over8}+\d^2{\pi\over4}+c(k)\d^{5/2}
+O(\d^3),
\label{asI}
\ee
where
$$
c(k)=-{8\over5}\int_0^\infty \,du{-ky+k^2+1\over[(y-k)^2+1]^{5/2}}y^{5/2}.
$$
\par
First three terms in this formula are calculated by standard way.
Let's show as we  can obtain the function $c(k)$. For this purpose the
following reception (\cite{Fed1}) is applicable. Let's calculate third
derivative with respect to $\d$ of the function $I(k,\d)$:
$$
{\pt^3I\over\pt\d^3}=-3\int_0^1dz\,\sqrt{(1-z)z}
{-kz+k^2\d+\d\over[(z-k\d)^2+\d]^{5/2}}.
$$
On the right hand side we replace $z$ by $\d y$ and we present the
integral as
\bb
{\pt^3I\over\pt\d^3}=-3\d^{-1/2}\int_0^{\infty}dy\,y^{5/2}
{-ky+k^2+1\over[(y-k)^2+1]^{5/2}}+O(1).
\label{de1}
\ee
\par
Solving the ordinary differential equation (\ref{de1}) in the
neighborhood of $\d=0$, we get:
$$
I(k,\d)=c_0+\d c_1+\d^2c_2+\d^{5/2}{8\over15}c_3(k)+O(\d^3),
$$
where
$$
C_3(k)=-3\int_0^\infty\,du{-ky+k^2+1\over[(y-k)^2+1]^{5/2}}y^{5/2}.
$$
After that it is easy to obtain the asymptotics (\ref{asI}).
\par
To define the value of the number $k$ we substitute the asymptotics
(\ref{a-b}) and (\ref{asI}) in (\ref{eq-E1}) and  equate to zero the
coefficients at identical powers of $\eta$. In the result we get at
$\eta^{5/4}$ the equation
$$
c(k)=0.
$$
\par
This is the transcendental equation for the definition of the
parameter $k$. The numerical solution gives $k\sim0.463$. Using the
formula (\ref{k1}), we get:
$$
\mu_1={k\over3}|\nu|,\quad \nu_1=\sqrt{{3\over2(3-k^2)}}.
$$
\par
To construct the asymptotics of $S(t)$ at $t\to t_*+0$ we use the
equation connecting the period of fast oscillations with its phase
(\cite{Fed}):
\bb
T=\sqrt{2}S'\int_\b^\a{dx\over\sqrt{(\a-x)(x-\b)[(x-m)^2+n^2]}}.
\label{eq-S2}
\ee
Present the integral in the right hand side as
$$
J={1\over\a-\b}\int_0^1{dz\over\sqrt{(1-z)z[(z-k\d)^2+\d]}}.
$$
After the same replacements, as at the construction of the asymptotics
${\pt^3I\over\pt\d^3}$, at $\d\to0$ we get:
$$
J={\d^{-1/2}\over\a-\b}\int_0^{\infty}{dy\over\sqrt{y[(y+k)^2+1]}}\,
+O(1).
$$
We substitute this expression in the equation (\ref{eq-S2}), use the
asymptotics $\d$ and $(\a-\b)$ at $\eta\to+0$ and in the result we
get:
$$
S'=(t-t_*)^{1/4}S_*(k)+O((t-t_*)^{1/2}),
$$
where
$$
S_*(k)={T\over\sqrt2}{2|m_*|^{1/2}\over C_*(k)}
\bigg({3\over6-2k^2}\bigg)^{1/4},
\quad C_*(k)=\int_0^\infty{dy\over\sqrt{y[(y-k)^2+1]}}.
$$
\par
The period of the oscillations for the function
$\stackrel{0}{U}(t_1,t)$ with respect to the variable $t_1$ in the
Krylov-Bogolubov's method is an arbitrary constant. Let's choose it
such, that $S_*(k)=1$:
\bb
T={S_*(k)\sqrt{2}C_*(k)\over2|u_*|^{1/2}}\bigg({3\over6-2k^2}\bigg)^{1/4}.
\label{T}
\ee
In the result the phase of the oscillations at $t\to t_*$ has form
\bb
S(t)={4\over5}(t-t_*)^{5/4}+O((t-t_*)^{3/2})+S_0,
\label{phase}
\ee
where $S_0$ is  some constant. Its value will be defined below at the
matching of the asymptotics (\ref{exp4}) and interior asymptotics
(\ref{exp2}), \ref{exp3}) at $t\to t_*+0$.
\par
Now we turn to the evaluation of the asymptotics for the function
$\stackrel{0}{U}$ at $t\to t_*+0$. Let's search for the asymptotics of
the function $\stackrel{0}{U}$ as
\bb
\stackrel{0}{U}(t_1,t)=u_*+W(\theta)+O((t-t_*)),
\label{exp5}
\ee
where $\theta=S(t)/\ve^{-1}.$
\par
Substitute this asymptotics to the equation for $\stackrel{0}{U}$
(\ref{eq-U}), in the result we get:
$$
(\pt_\theta W)^2=-W^4+4u_*W^3+O((t-t_*)).
$$
It is easy to see, that in the leading order we have the function
$\stackrel{0}{w}(\theta)$, defining the second interior asymptotic
expansion.
\par
The formula (\ref{exp5}) is suitable at $|W(\theta)|\gg|t-t_*|$. When
$W(\theta)$ is small, we consider other asymptotic formula for the
function $\stackrel{0}{U}(t_1,t)$:
\bb
\stackrel{0}{U}(t_1,t)=u_*+\sqrt{t-t_*}p(S(t))+O((t-t_*)^2).
\label{exp6}
\ee
Substitute this formula to the second-order equation for the function
$\stackrel{0}{U}(t_1,t)$ (\ref{eq-2U}), in the result we get:
$$
p''+6u_*p^2+u_*=0.
$$
This equation coincides with the equation for the asymptotics of the
first correction of the first interior asymptotic expansion. The
boundary conditions for the function $p(S(t))$ is obtained from the
condition of the matching (\ref{exp6}) with the asymptotics of the
expansion (\ref{exp5}) at $|\theta|\to\infty$. The phase shift $S_0$
in the formula (\ref{phase}) is finally defined at the matching of the
asymptotic expansions (\ref{exp5}), (\ref {exp6}) with asymptotics of
the interior asymptotic expansions. This get: $S_0=0$.
\par
I am grateful to A.N. Belogrudov, S.G. Glebov, L.A. Kalykin, V.Yu.
Novokshenov and B.I. Suleimanov for fruitful discussions and also V.E.
Adler for the help in the realization of the numerical calculating.
\par
This work was supported by Russian Foundation for Basic Research
97-01-00459.

\end{document}